\documentclass[12pt]{article}
\usepackage{latexsym}
\usepackage{epsfig}

\usepackage{graphicx}
\usepackage{epstopdf}

\usepackage{epsfig,amssymb,amsmath,amscd}

\newcommand{\bq}{\begin{equation}}
\newcommand{\eq}{\end{equation}}
\newcommand{\bqa}{\begin{eqnarray}}
\newcommand{\eqa}{\end{eqnarray}}
\newcommand{\ben}{\begin{enumerate}}
\newcommand{\een}{\end{enumerate}}
\newcommand{\bc}{\begin{center}}
\newcommand{\ec}{\end{center}}
\newcommand{\bqb}{\begin{eqnarray*}}
\newcommand{\eqb}{\end{eqnarray*}}

\def\pr#1#2#3{Phys. Rev. ${\bf{#1}}$, #2 (#3)}

\def\pl#1#2#3{Phys. Lett. ${\bf{#1}}$, #2 (#3)}

\def\np#1#2#3{Nucl. Phys. ${\bf{#1}}$, #2 (#3)}

\def\jhep#1#2#3{JHEP ${\bf{#1}}$, #2 (#3)}
\def\epj#1#2#3{Eur. Phys. J. ${\bf{#1}}$, #2 (#3)}

\def\jmp#1#2#3{J. Mod. Phys. ${\bf{#1}}$, #2 (#3)}

\begin{document}
\pagenumbering{arabic}
\thispagestyle{empty}
\def\thefootnote{\fnsymbol{footnote}}
\setcounter{footnote}{1}

%\vspace{2cm}
\begin{flushright}
Jan. 31, 2018\\
%arXiv: \\
 \end{flushright}

\begin{center}
{\Large {\bf Dark matter interaction between
massive standard particles.}}\\
 \vspace{1cm}
%-----------------------------------------------------------------
{\large F.M. Renard}\\
\vspace{0.2cm}
Laboratoire Univers et Particules de Montpellier,
UMR 5299\\
Universit\'{e} Montpellier II, Place Eug\`{e}ne Bataillon CC072\\
 F-34095 Montpellier Cedex 5, France.\\
\end{center}

\vspace*{1.cm}
\begin{center}
{\bf Abstract}
\end{center}

We propose further tests of the assumption that the mass of the 
heavy standard particles ($Z,W,t,...$) arises from a special coupling 
with dark matter. We look for effects of new interactions due to dark
matter exchanges between heavy particles in several $e^+e^-$ and 
hadronic collision processes.

\vspace{0.5cm}

\def\thefootnote{\arabic{footnote}}
\setcounter{footnote}{0}
\clearpage

\section{INTRODUCTION}

In a previous paper (\cite{DMmass}) we have supposed that heavy standard particles
(for example $Z,W,t,...$) get their mass from the dark matter (DM) environment.
The connection between heavy SM particles and DM 
(whose status is for example reviewed in \cite{rev1}) could occur through
mediators (scalar, vector ,..., see \cite{rev1,Portal}), or through a yet undefined
more complex way.
In our work it was described by effective couplings (for example $Z-Z-DM$)
which may then also generate dark matter production in association with heavy particles.
We have given illustrations for $e^+e^-\to Z+X,~W^+W^-+X, ~t\bar t+X$ where $X$
represents invisible multiparticle DM.\\
This picture has some similarity with the hadronic case where hadrons get
their mass from the strong quark binding interaction and where multiparticle
production is automatically generated through the parton model description.\\

In the present paper we explore another consequence of our assumption of the existence
of a special connection between heavy SM particles and DM. It would consist in
the presence of a new interaction between heavy SM particles resulting from
the exchange of this invisible set of multiparticle DM. 
This may be again somewhat similar to the hadronic case where strong 
final state interactions appear between hadrons 
after their production through electroweak processes.\\
This can appear through an involved mechanism and not simply through a 
mediator (for example a Higgs boson) exchange. Our aim is only
to show what type of effects could appear that way and which specific relations may exist
between the various processes involving heavy SM particles.\\ 
We will consider the production of pairs of heavy particles, $t\bar t$, $ZZ$, $W^+W^-$  in $e^+e^-$, $\gamma\gamma$ and hadronic collisions.\\
Without a precise model for these new DM interactions we will make 
illustrations using an arbitrary effective description which protects the SM prediction
at low energy (see the CSM concept \cite{CSMrev}).\\
A strategy for the determination of the actual interaction between
heavy pairs will be proposed by comparing its effects in the various processes.\\
Contents: Presentation of the effective interaction in Section 2; applications to several processes in Section 3. 
Further developments and other possible interpretations of new interactions among
heavy particles will be discussed in the conclusion.\\

\section{EFFECTIVE DM INTERACTION}

Following the assumption presented in the introduction 
we will consider possible DM interactions between $t\bar t$, $ZZ$, $WW$ pairs.
This could appear either after their standard production from
light states (not submitted to DM interactions) like final state interaction 
in hadronic production or during their production from heavy states (the new DM interaction occuring in the initial state, in the final state 
or in the $t,u$ exchange channels).\\

In the absence of a specific model for the description of the DM interaction
we will use an effectice form allowing the computation (not necessarily perturbative) of its initial, final or exchange effects.\\

In the simplest case with only an s-channel final state interaction in a 2 body
production of heavy particles ($AB$) from initial light particles insensitive
to DM interactions, we write the corresponding amplitude as

\bq
R=R_{SM}[1+F_{AB}(s)]  ~~, \label{R}
\eq
In this first study we do not specify the helicity amplitudes of the $AB$
state. In a more involved analysis we may use separate functions for each helicity combination.\\

The simplest examples, shown in Fig.1, are $e^+e^- \to t\bar t$,
$gg \to t\bar t$ (from which one can deduce the $\gamma\gamma \to t\bar t$ case by suppressing the second Born diagram and replacing the gluon coupling by the photon one), 
$e^+e^- \to ZZ$ and $e^+e^- \to W^+W^-$.
The corresponding DM interactions in the $t\bar t$, $ZZ$ or $WW$  final states 
only appear in the $s$ channel and are described by the  $F_{t\bar t}(s)$,
$F_{ZZ}(s)$ and $F_{WW}(s)$ functions.\\

We then consider the richer processes $ZZ\to ZZ$ (Fig.2), 
$WW\to WW$ (Fig.3), $ZZ\to W^+W^-$ (Fig.4),
$ZZ\to t\bar t$ (Fig.5) and $W^+W^-\to t\bar t$ (Fig.6).
As we can see in the corresponding figures they involve DM interactions 
between heavy pairs in $s$, $t$ and $u$ channels.
Precisely, for the respective processes, we write 

\bq
ZZ\to ZZ~~~~~~~~R=R_{SM}
[1+2F_{ZZ}(s)+2F_{ZZ}(t)+2F_{ZZ}(u)] ~~, \label{FZZ} 
\eq
\bq
W^+W^-\to W^+W^- ~~~~~~~~R=R_{SM}
[1+2F_{WW}(s)+2F_{WW}(t)+2F_{WW}(u)] ~~, \label{FWW} 
\eq
\bq
ZZ\to W^+W^-~~~~~~~~R=R_{SM}
[1+F_{ZZ}(s)+F_{WW}(s)+2F_{ZW}(t)+2F_{ZW}(u)] ~~, \label{FZW} 
\eq
\bq
ZZ\to t\bar t~~~~~~~~R=R_{SM}
[1+F_{ZZ}(s)+F_{t\bar t}(s)+2F_{Zt}(t)+2F_{Zt}(u)] ~~, \label{FZt} 
\eq
\bq
W^+W^-\to t\bar t~~~~~~~~R=R_{SM}
[1+F_{WW}(s)+F_{t\bar t}(s)+2F_{Wt}(t)+2F_{Wt}(u)] ~~, \label{FWt} 
\eq

A possible simplification could arise (especially at high energy) from the structures of the diagrams of Fig.1-6

\bq
F^2_{AB}(x)=F_{AA}(x)F_{BB}(x)~~, \label{FAB}
\eq
\noindent
so that the only unknowns are $F_{t\bar t}(x)$,
$F_{ZZ}(x)$ and $F_{WW}(x)$ for $x=s,t,u$.\\

A strategy for the search and for the analysis of such DM effects
could be the following one:\\

---- ~~ First the $F_{t\bar t}(s)$ function
could be determined in $e^+e^- \to t\bar t$, 
and/or in $\gamma\gamma \to t\bar t$ and/or in $gg \to t\bar t$, these different processes
allowing to make consistency checks.\\

---- ~~ Separately  $F_{ZZ}(s)$ could be determined in $e^+e^- \to ZZ$ and checked 
in the $t,u$ channels with $ZZ \to ZZ$.\\

----~~~ One would also obtain $F_{WW}(s)$ in $e^+e^- \to W^+W^-$ and/or also in 
$\gamma\gamma \to W^+W^-$ and check it in the $t,u$ channels with $W^+W^-\to W^+W^-$.\\

---- ~~ Then, mixed effects, involving different heavy pairs, in $ZZ \to W^+W^-$,
$ZZ \to t\bar t$ and $W^+W^- \to t\bar t$ should globally confirm the DM assumption and its characteristics obtained from the studies of the preceeding simple processes.\\

In the next section we give simple illustrations of these possibilities using 
trial forms for the effective functions.
We will use a simple form inspired by the one appearing at high energy from
usual triangle and box diagrams  (\cite{logs}) where, 
depending on the spin and couplings of the exchanged particles, $log$ or $log^2$
terms may appear. In our cases the exchange may contain the basic Higgs boson, 
possible excited states and more complex DM sets.
For simplicity we use a single logarithmic energy dependence 
with a scale obtained by imposing the normalization to the SM value at threshold 
(this could be motivated by the CSM concept \cite{CSMrev}).
\bq
F(s)=c~ln{-s\over s_{th}} ~~, \label{Fs}
\eq
where $c$ is for the moment an unknown strength coefficient
and $s_{th}$ corresponds to the threshold
for production of the pair of heavy particles.
A priori one could expect that the sizes of the $F_{t\bar t}(x)$, $F_{ZZ}(x)$ 
and $F_{WW}(x)$ functions would be somewhat different.
They may be related to the masses, $m_t$, $m_Z$ and $m_W$, if the involved DM interaction is the one which creates these masses (\cite{DMmass}). In the illustrations
we will use $c=m^2_{t,Z,W}/m^2_0$ with $m_0=0.5$ TeV. 
But we repeat that these choices are totally arbitrary. They are only made in order
to easily do computations in the various channels and to clearly illustrate a strategy for
the determination of the structure of the assumed DM interaction.\\

\section{PRECISE APPLICATIONS}

\subsection{Simple processes $e^+e^- \to t\bar t$, $gg \to t\bar t$, $e^+e^- \to ZZ$, $e^+e^- \to W^+W^-$}

Following Fig.1 the usual Born diagram would be corrected by a final state
DM interaction between the top quarks or $ZZ$ or $W^+W^-$, according to eqs.(\ref{R}),(\ref{Fs})
with a function $F_{t\bar t}(s)$, $F_{ZZ}(s)$ 
or $F_{WW}(s)$. In the illustration of Fig.7 
the corresponding energy dependence of the cross section is shown 
for $\theta=\pi/3$ with $s_{th}=4m^2_{t,Z,W}$ and $c=m^2_{t,Z,W}/m^2_0$ . No modification of the shape of the angular distribution is expected 
from such a final state interaction.\\
The $\gamma\gamma \to t\bar t$ case can be deduced from $gg \to t\bar t$
keeping only the first $gg \to t\bar t$ Born diagram (with its symmetrical term);
with the same $F_{t\bar t}(s)$  one obtains similar effects\\
As mentioned in the previous sections the illustrations correspond to
an arbitrary choice only used for presenting a
global strategy of analyzing possible DM interactions 
among heavy particles. In practice a fit of the experimental results would
determine the presently unknown $F_{t\bar t}(s)$, $F_{ZZ}(s)$ 
$F_{WW}(s)$functions.\\

\subsection{$ZZ \to ZZ$}

This is a more complex process which only involves $ZZ$
DM interaction but which would occur simultaneously in $s$, $t$ and $u$ channels
as one can see in Fig.2 with the corresponding Born diagrams (with $ZZ$
symmetrization). The functions $F_{ZZ}(s)$, $F_{ZZ}(t)$, $F_{ZZ}(u)$
will now appear with

\bq
R=R_{SM}
[1+2F_{ZZ}(s)+2F_{ZZ}(t)+2F_{ZZ}(u)] ~~, \label{FZZ} 
\eq

The results are shown in Fig.8 for the energy and the (symmetrical)
angular dependences.\\
Effects are obviously larger then in the case of simple processes when 
using the same functions.\\

\subsection{$W^+W^-\to W^+W^-$}

Similarly this process will involve the $WW$
DM interaction in $s$, $t$ and $u$ channels
as one can see in Fig.3 with the corresponding Born diagrams. 
The functions $F_{WW}(s)$, $F_{WW}(t)$, $F_{WW}(u)$ will also appear with
\bq
R=R_{SM}[1+2F_{WW}(s)+2F_{WW}(t)+2F_{WW}(u)] 
\eq

The results are shown in Fig.9  for the energy and the (forward peaked)
angular dependences.\\

\subsection{$ZZ \to W^+W^-$}

This is a new case whose diagrams are drawn in Fig.4 and which involves
simple $F_{ZZ}(s)$, $F_{WW}(s)$ and mixed $F_{ZW}(t)$, $F_{ZW}(u)$ functions
(with $x_{th}=(m_Z+m_W)^2$, $x=t,u$); using (\ref{FAB}) we write

\bqa
R&=&R_{SM}[1+F_{ZZ}(s)+F_{WW}(s)+2F_{ZW}(t)+2F_{ZW}(u)]\nonumber\\
&&\simeq R_{SM}[1+F_{ZZ}(s)+F_{WW}(s)+F_{ZZ}(t)+F_{WW}(t)+F_{ZZ}(u)+F_{WW}(u)]
\eqa
 
The results are shown in Fig.10 for the energy and the (symmetrical)
angular dependences.\\

\subsection{$ZZ \to t\bar t$}

This is another type of new case (see Fig.5) involving simple 
$F_{ZZ}(s)$, $F_{t\bar t}(s)$ and mixed $F_{Zt}(t)$, $F_{Zt}(u)$ functions
(with $x_{th}=(m_Z+m_t)^2$); using (\ref{FAB})

\bqa
R&=&R_{SM}[1+F_{ZZ}(s)+F_{WW}(s)+2F_{ZW}(t)+2F_{ZW}(u)] \nonumber\\ 
&&\simeq R_{SM}[1+F_{ZZ}(s)+F_{t\bar t}(s)+F_{ZZ}(t)+F_{t\bar t}(t)+F_{ZZ}(u)+F_{t\bar t}(u)]
\eqa

The results are shown in Fig.11 for the energy and the (symmetrical)
angular dependences.\\

\subsection{$W^+W^- \to t\bar t$}

Finally we consider the similar process (see Fig.6) involving simple 
 $F_{WW}(s)$, $F_{t\bar t}(s)$ and mixed $F_{Wt}(t)$, $F_{Wt}(u)$ functions
(with $x_{th}=(m_W+m_t)^2$), using (\ref{FAB})

\bqa
R&=&R_{SM}[1+F_{WW}(s)+F_{t\bar t}(s)+2F_{Wt}(t)+2F_{Wt}(u)]\nonumber\\ 
&&\simeq R_{SM}[[1+F_{WW}(s)+F_{t\bar t}(s)+F_{WW}(t)+F_{t\bar t}(t)+F_{WW}(u)+F_{t\bar t}(u) ]
\eqa

The results are shown in Fig.12  for the energy and the (forward peaked)
angular dependences.\\

We stop there our illustrations but obviously other processes could
be considered with for example different initial gauge bosons or quarks
and possibly with production of Higgs bosons.\\

As mentioned in the previous section and at the begining of this section,
after a first experimental analysis devoted to the simple processes and
having produced fits for $F_{t\bar t},F_{ZZ},F_{WW}$,
the illustrations in Fig.8-12 should be replaced by new ones obtained by 
using these fitted functions instead of the trial ones based on eq.(\ref{Fs}).\\

\section{CONCLUSION}

Assuming that heavy standard particles may be sensitive to additional interactions
related to the DM environment we have shown how this would affect the SM predictions for
several 2 body processes. In the simplest processes where heavy particles are
produced by light ones the situation would be rather similar to the 
case of final state interactions in hadronic production through electroweak processes.\\
In this spirit we have illustrated possible effects of a DM interaction
in the production of pairs of heavy particles ($Z,W,t,...$) in $e^+e^-$, 
$\gamma\gamma$ and hadronic collisions by modifying the SM amplitudes
with effective forms $F_{AB}(x),x=s,t,u$.\\
The comparison of these various processes should be instructive about the
presence and the structure of such  possible departures from SM expectations.
With that aim we have proposed a strategy for the required analyses
of the various processes:\\
1) determination of the basic $F_{AB}$ functions using
the simplest processes with initial light (non DM interacting) states,\\
2) checks of their adequacy, of their structure, in particular of their
factorization rule, eq.(\ref{FAB}), in processes with initial heavy particles.\\
The experimental possibilities at present and future colliders can be
found for example for $e^+e^-$ in \cite{Moortgat, Craig},
for photon-photon collisions in \cite{gammagamma}, and for hadronic collisioins
in \cite{Contino, Richard}.\\
Further developments of our proposal could concern the following phenomenological and
theoretical points. 
Precise analyses should be done at the level of the helicity amplitudes
with polarization measurements.
For a better accuracy in the analyses of departures
due to DM interactions higher order corrections, 
and not only Born amplitudes, will be required.
Possible applications to more complex (3 body, 4 body, ...) processes
may be considered.\\
Obviously there are fundamental theoretical points to examine like
the modelization of the DM interaction between heavy particles and
the role of the Higgs boson (occuring as an internal or an external contribution)
in this DM exchange.\\
There are also other effects which may interfer with those of the
yet undefined DM interactions, for example the possibility of
compositeness \cite{comp}, especially
top quark and Higgs boson compositeness (affecting also 
the longitudinal gauge bosons)
\cite{Hcomp2,Hcomp3,Hcomp4,partialcomp,Tait} which may create competitive anomalous
effects among heavy particles (\cite{CSMrev,trcomp,ttincl}).
For example a compositeness form factor and a DM final state interaction
may be largely competitive.
Detailed studies of possible ways to identify the origin of such
anomalous effects and to differentiate DM from compositeness should be done.\\

\newpage
\begin{figure}[p]
\vspace{-0cm}
\[
\hspace{-2cm}\epsfig{file=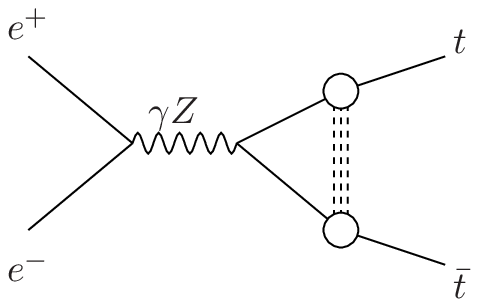 , height=3.cm}
\]\\
\vspace{-0cm}
\[
\hspace{-2cm}\epsfig{file=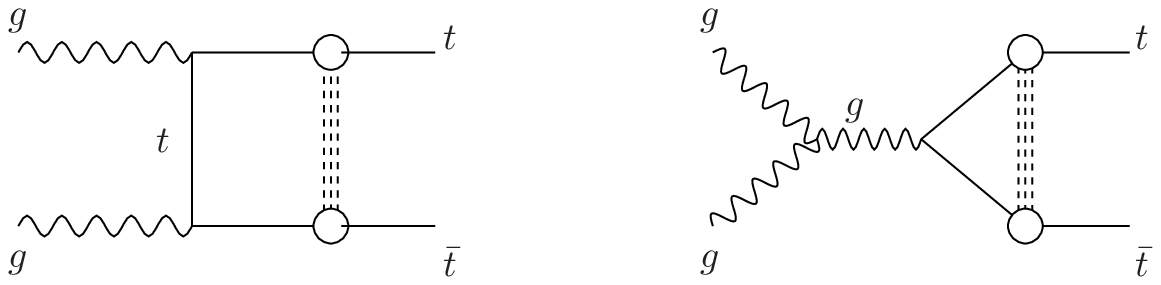 , height=3.cm}
\]\\
\vspace{-0cm}
\[
\hspace{-2cm}\epsfig{file=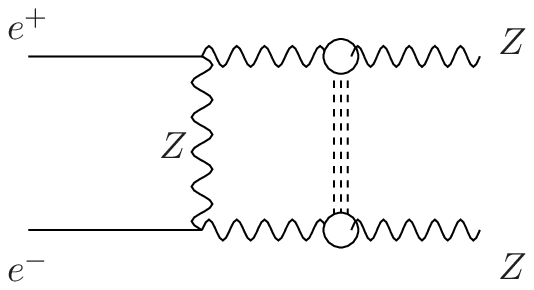 , height=3.cm}
\]\\
\vspace{-0cm}
\[
\hspace{-2cm}\epsfig{file=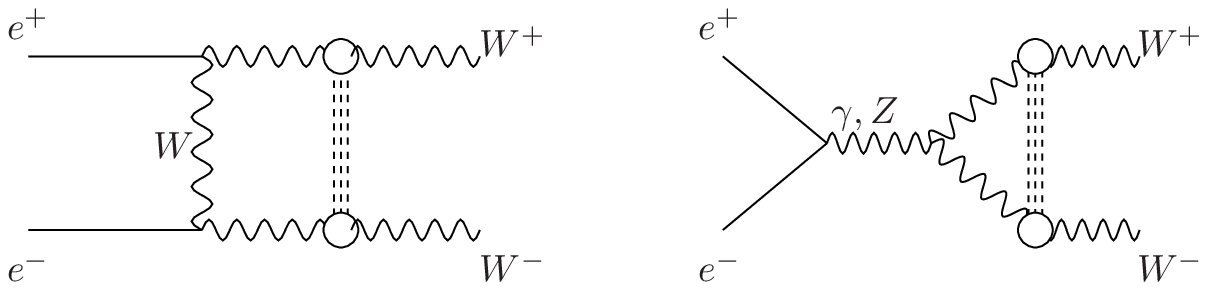 , height=3.cm}
\]\\
\vspace{-0cm}
\caption[1] {Simple processes with production of a pair of
massive particles submitted to final state DM interaction; $ZZ$ and $gg$ symmetrizations are applied.}
\end{figure}

\clearpage

\begin{figure}[p]
\vspace{-0cm}
\[
\hspace{-2cm}\epsfig{file=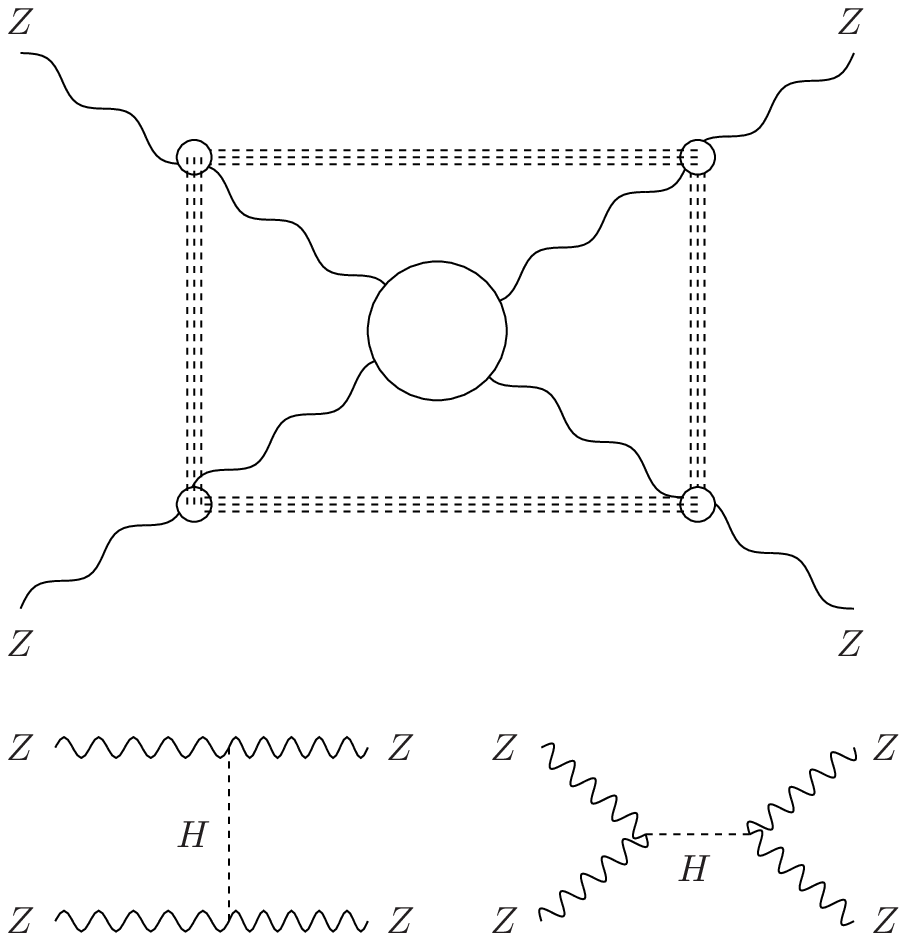 , height=10.cm}
\]\\
\vspace{-0cm}
\caption[1] {The process $ZZ\to ZZ$ with DM interaction in
both sides of the $s,t$ and $u$ channels. The center loop refers to the Born diagrams
drawn at the lower level; $ZZ$ symmetrization is applied.}
\end{figure}

\clearpage
\begin{figure}[p]
\vspace{-10cm}
\[
\hspace{-1cm}\epsfig{file=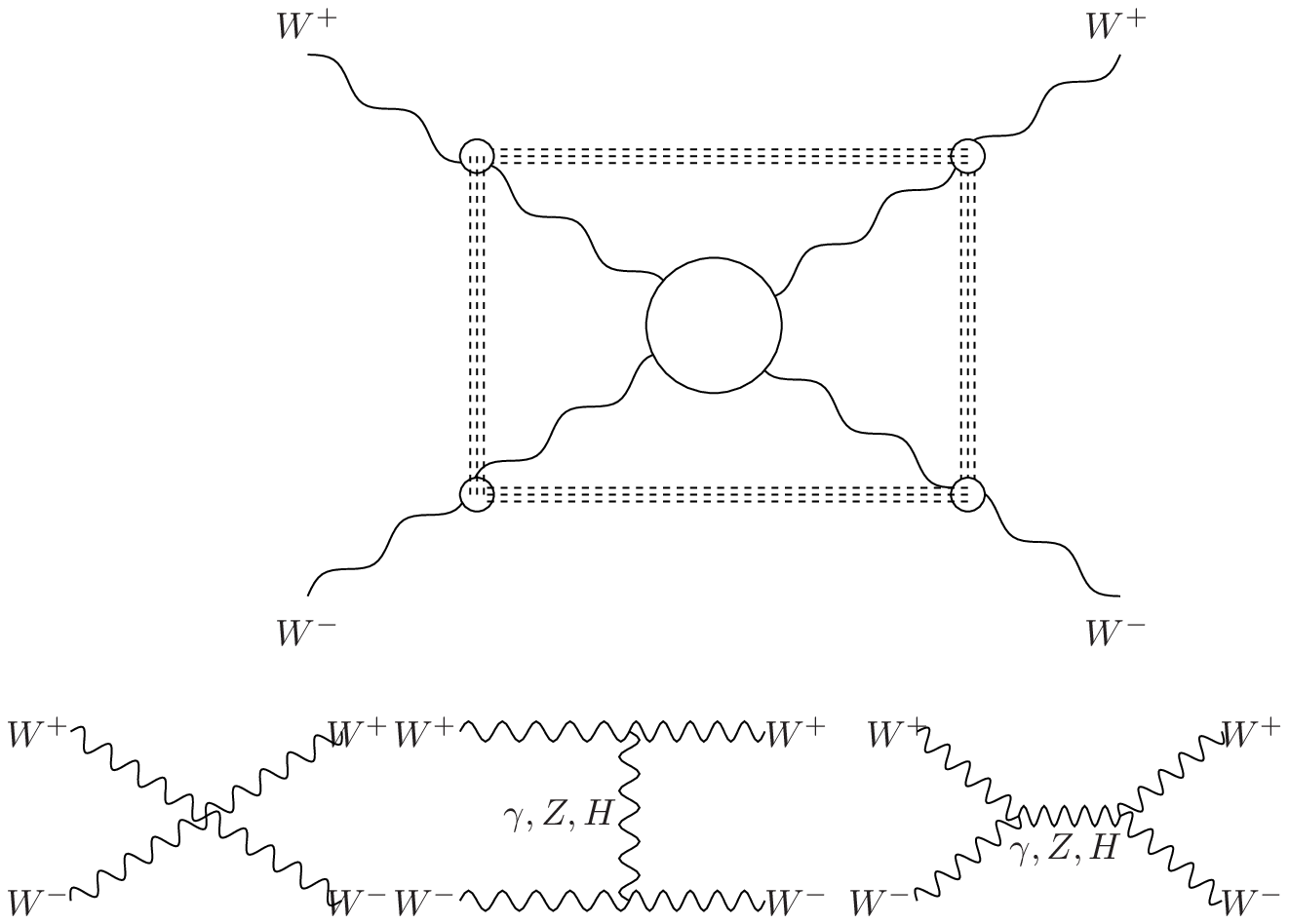 , height=30.cm}
\]\\
\vspace{-13cm}
\caption[1] {The process $W^+W^-\to W^+W^-$ with DM interaction in
both sides of the $s,t$ and $u$ channels. The center loop refers to the Born diagrams
drawn at the lower level.}
\end{figure}

\clearpage

\begin{figure}[p]
\vspace{-10cm}
\[
\hspace{-1cm}\epsfig{file=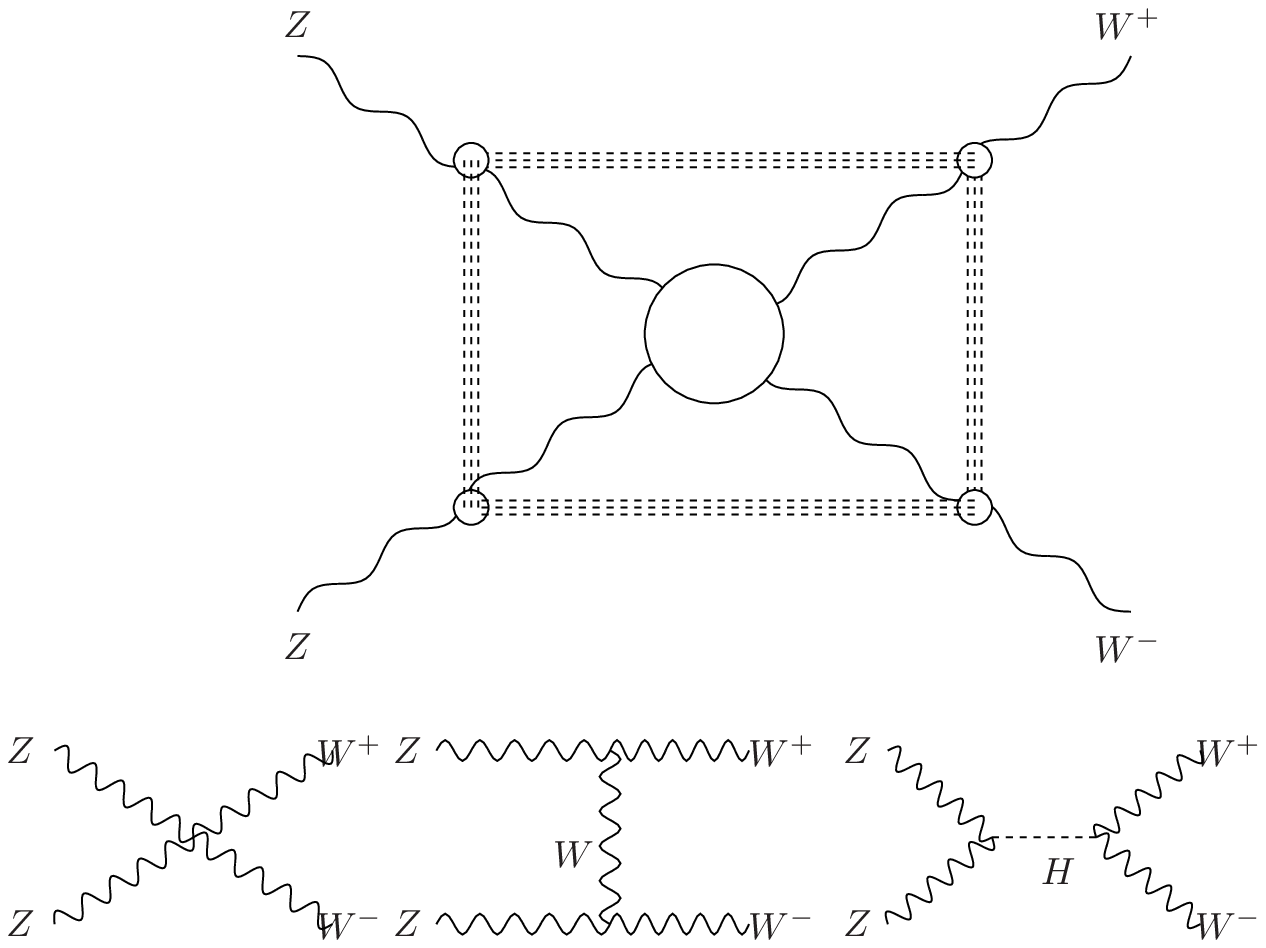 , height=30.cm}
\]\\
\vspace{-15cm}
\caption[1] {The process $ZZ\to W^+W^-$ with DM interaction in
both sides of the $s,t$ and $u$ channels. The center loop refers to the Born diagrams
drawn at the lower level; $ZZ$ symmetrization is applied.}
\end{figure}

\clearpage

\begin{figure}[p]
\vspace{-0cm}
\[
\hspace{-2cm}\epsfig{file=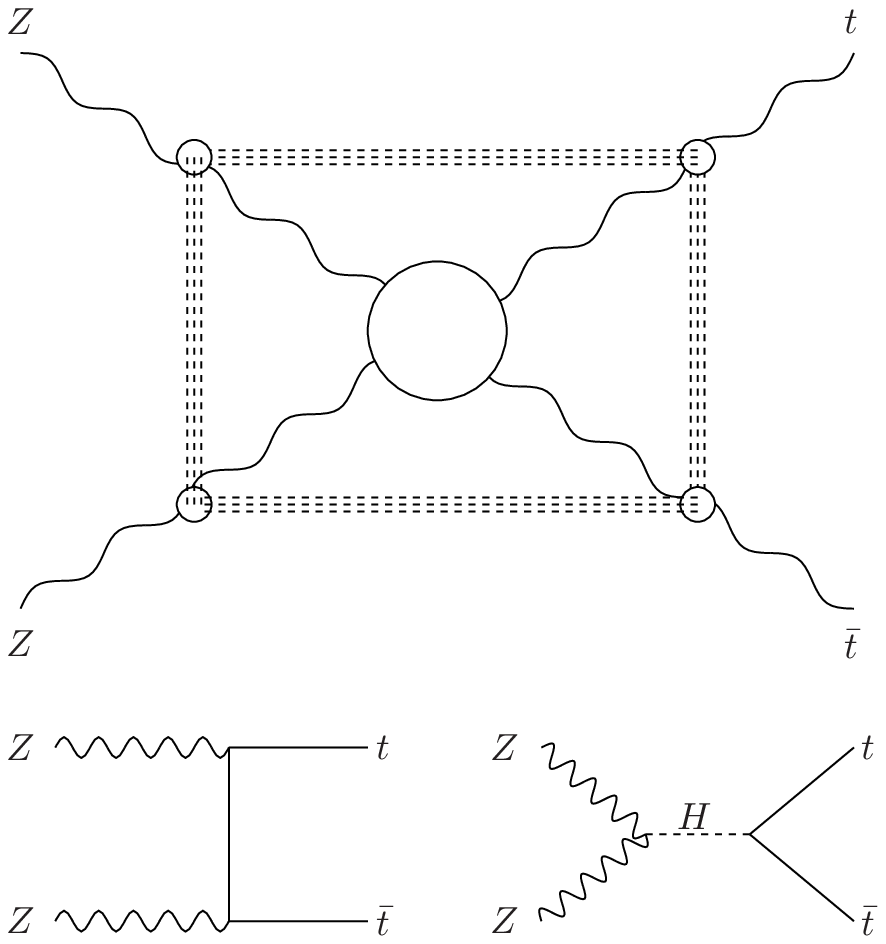 , height=30.cm}
\]\\
\vspace{-15cm}
\caption[1] {The process $ZZ\to t\bar t$ with DM interaction in
both sides of the $s,t$ and $u$ channels. The center loop refers to the Born diagrams
drawn at the lower level; $ZZ$ symmetrization is applied.}
\end{figure}

\clearpage

\begin{figure}[p]
\vspace{-0cm}
\[
\hspace{-2cm}\epsfig{file=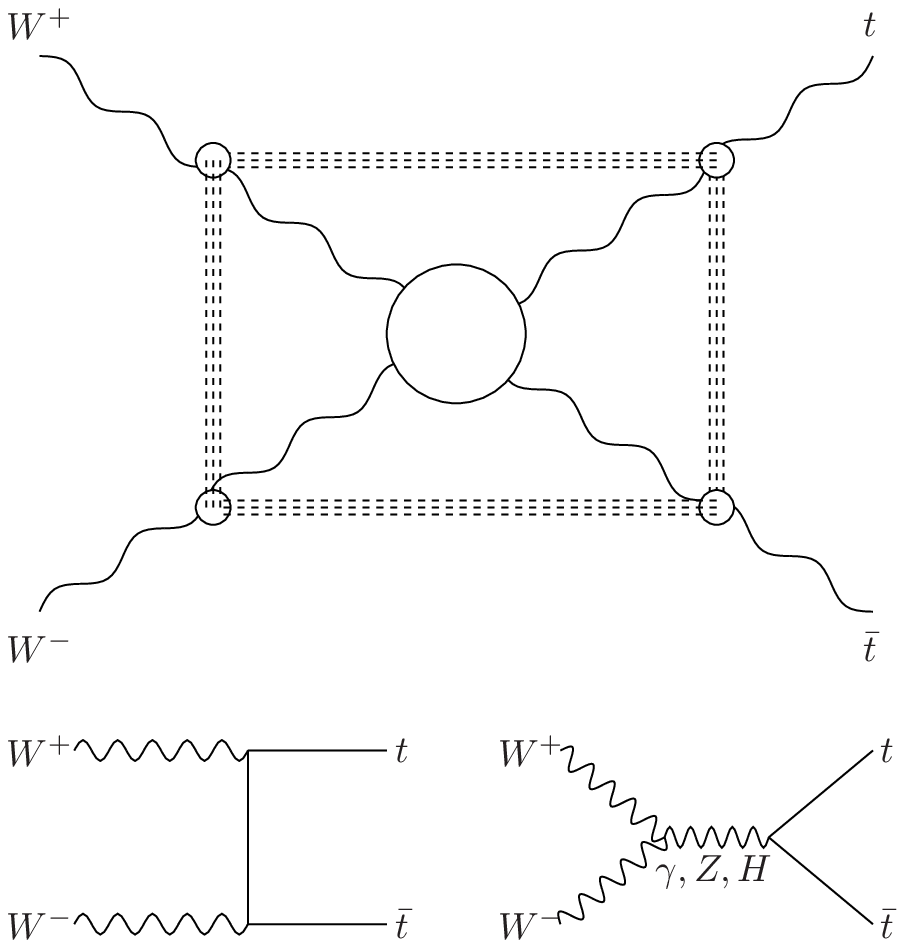 , height=30.cm}
\]\\
\vspace{-15cm}
\caption[1] {The process $W^+W^-\to t\bar t$ with DM interaction in
both sides of the $s,t$ and $u$ channels. The center loop refers to the Born diagrams
drawn at the lower level.}
\end{figure}

\clearpage

\begin{figure}[p]
\[
\epsfig{file=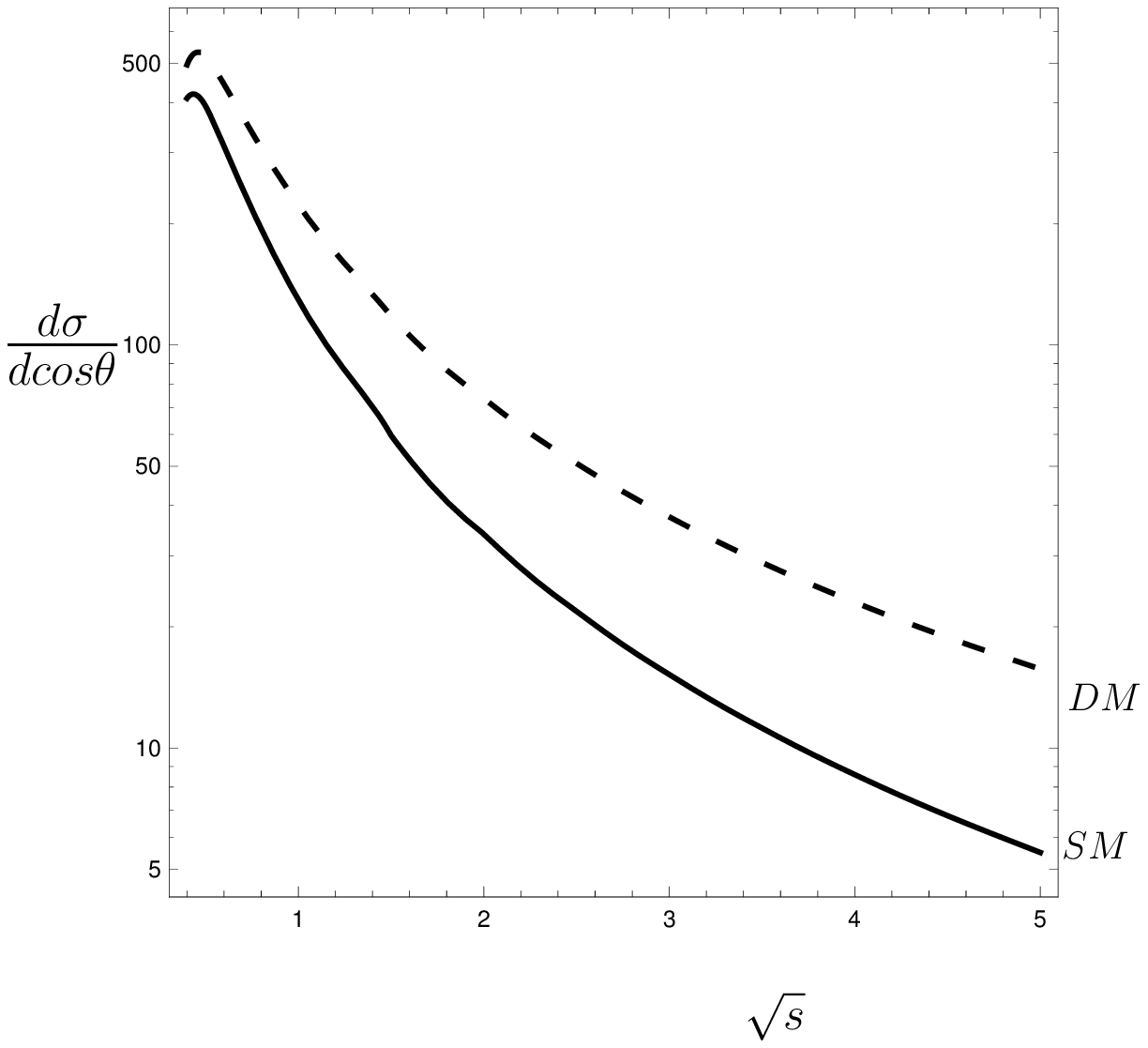 , height=17.cm}\hspace{-5cm}
\epsfig{file=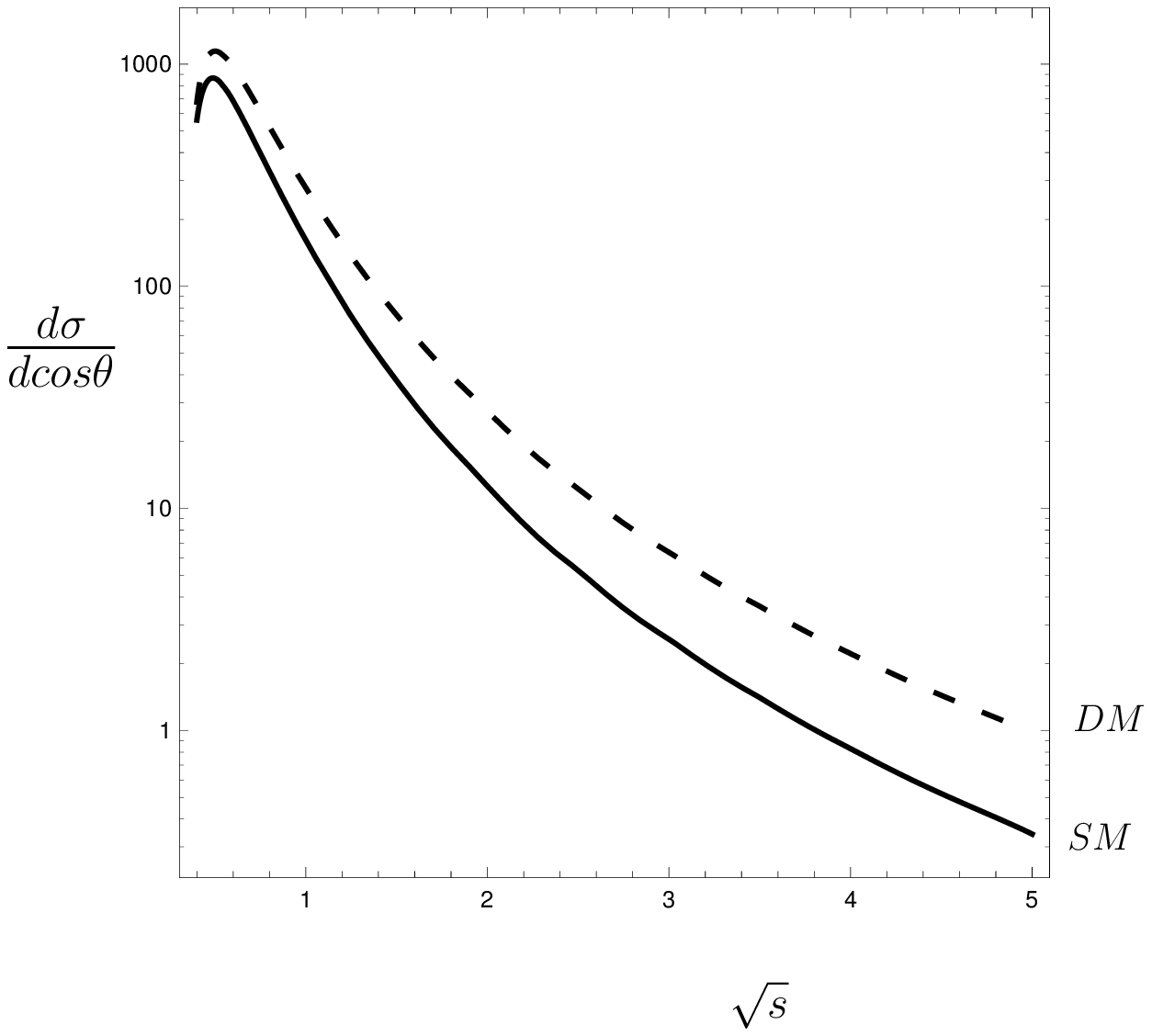 , height=17.cm}
\]
\\
\vspace{-10cm}
\[
\epsfig{file=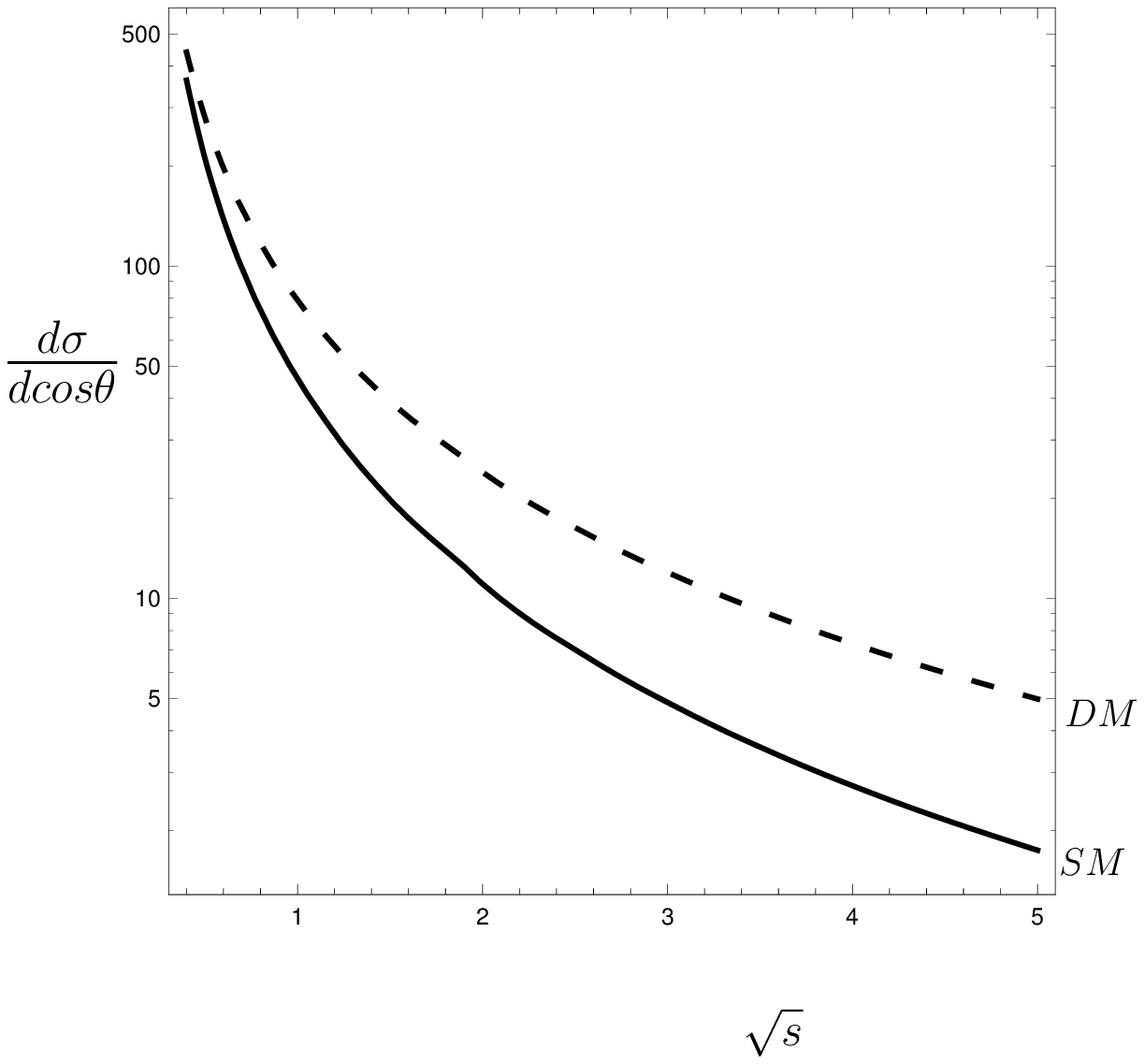 , height=17.cm}\hspace{-5cm}
\epsfig{file=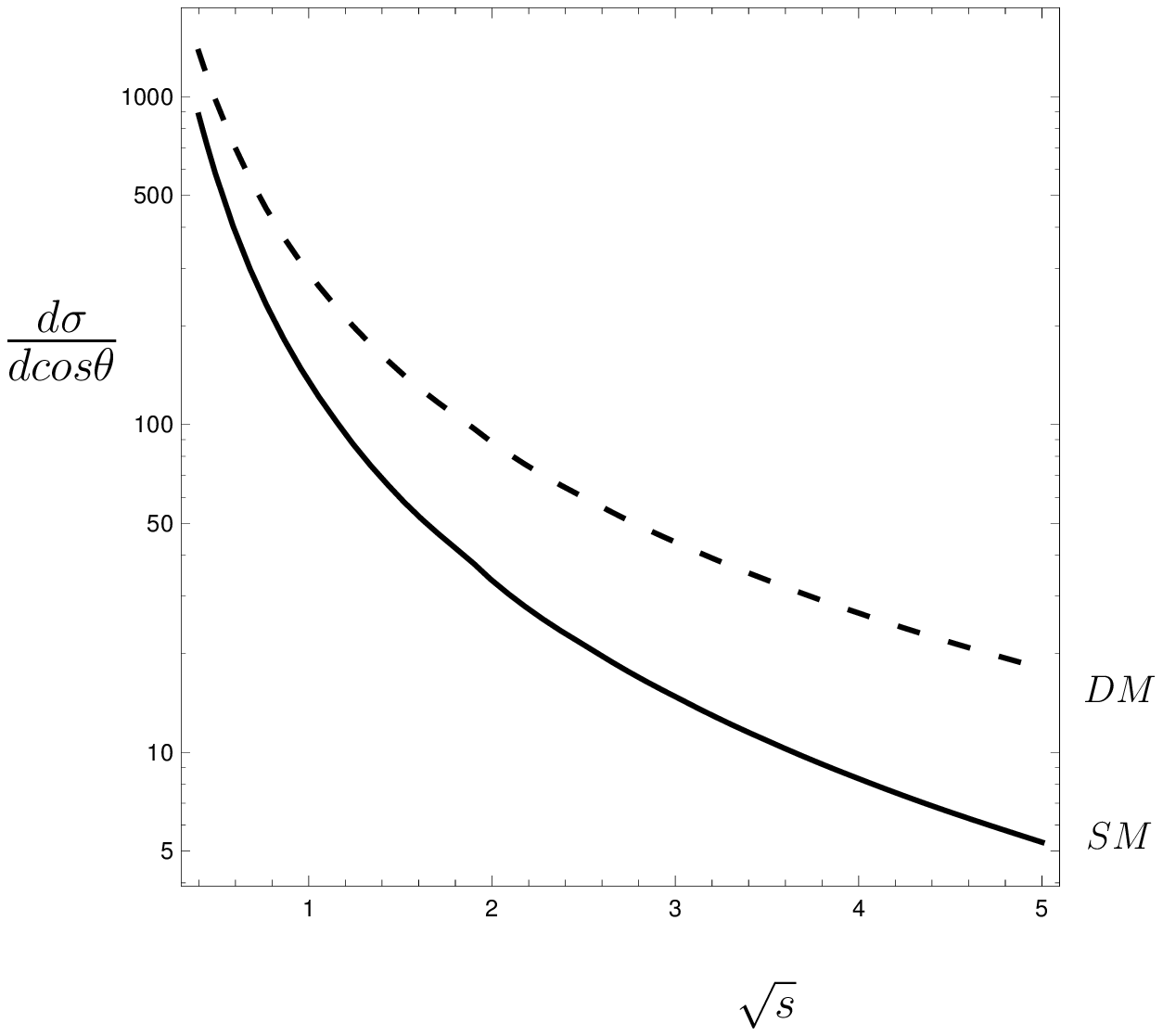 , height=17.cm}
\]
\\
\vspace{-7cm}
\caption[1] {DM effects in $e^+e^-\to t\bar t$, $gg \to t\bar t$ (upper level
and $e^+e^-\to W^+W^-$, $e^+e^-\to ZZ$ (lower level).}
\end{figure}

\clearpage

\begin{figure}[p]
\[
\epsfig{file=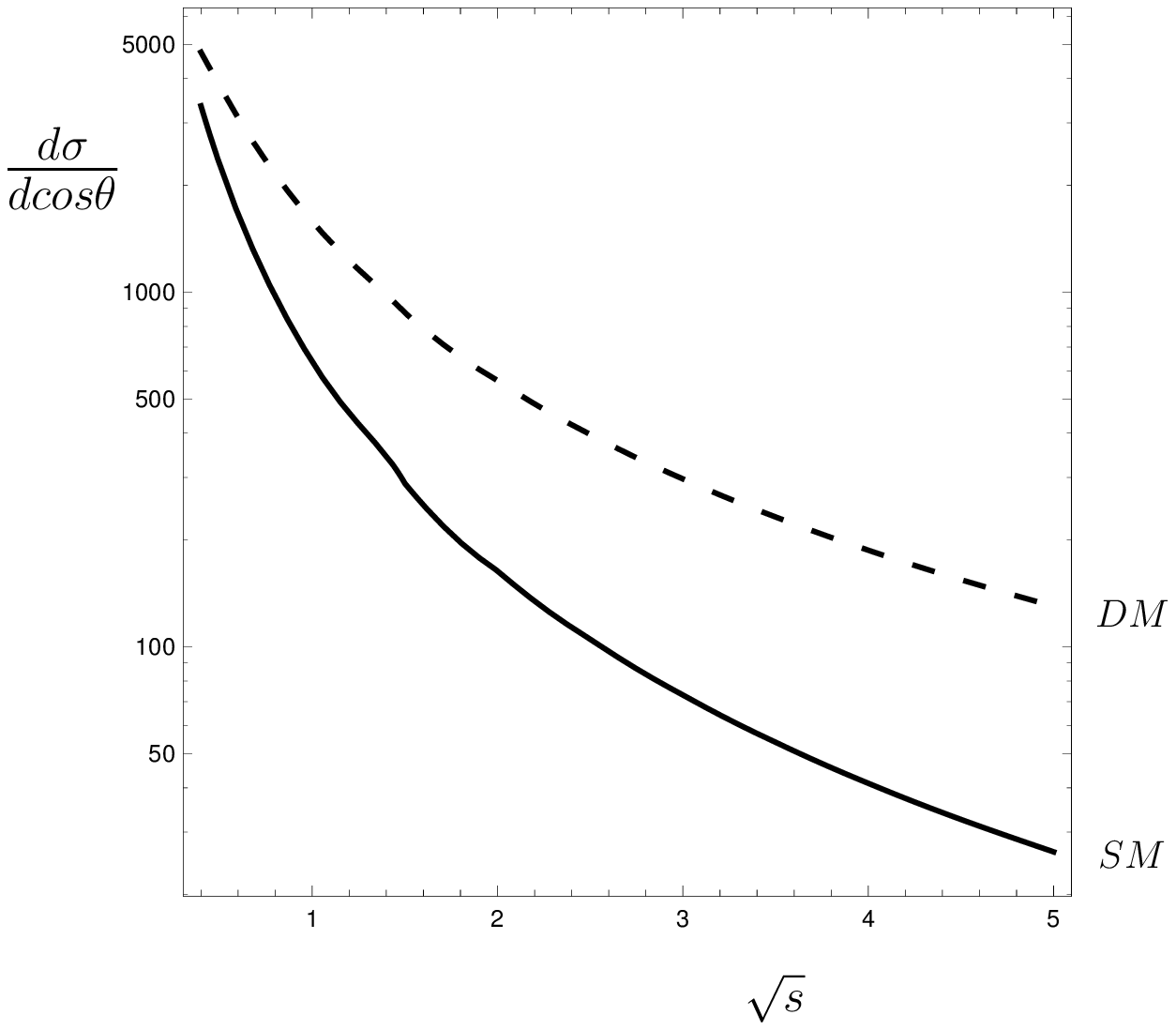 , height=20.cm}
\]\\
\vspace{-12cm}
\[
\epsfig{file=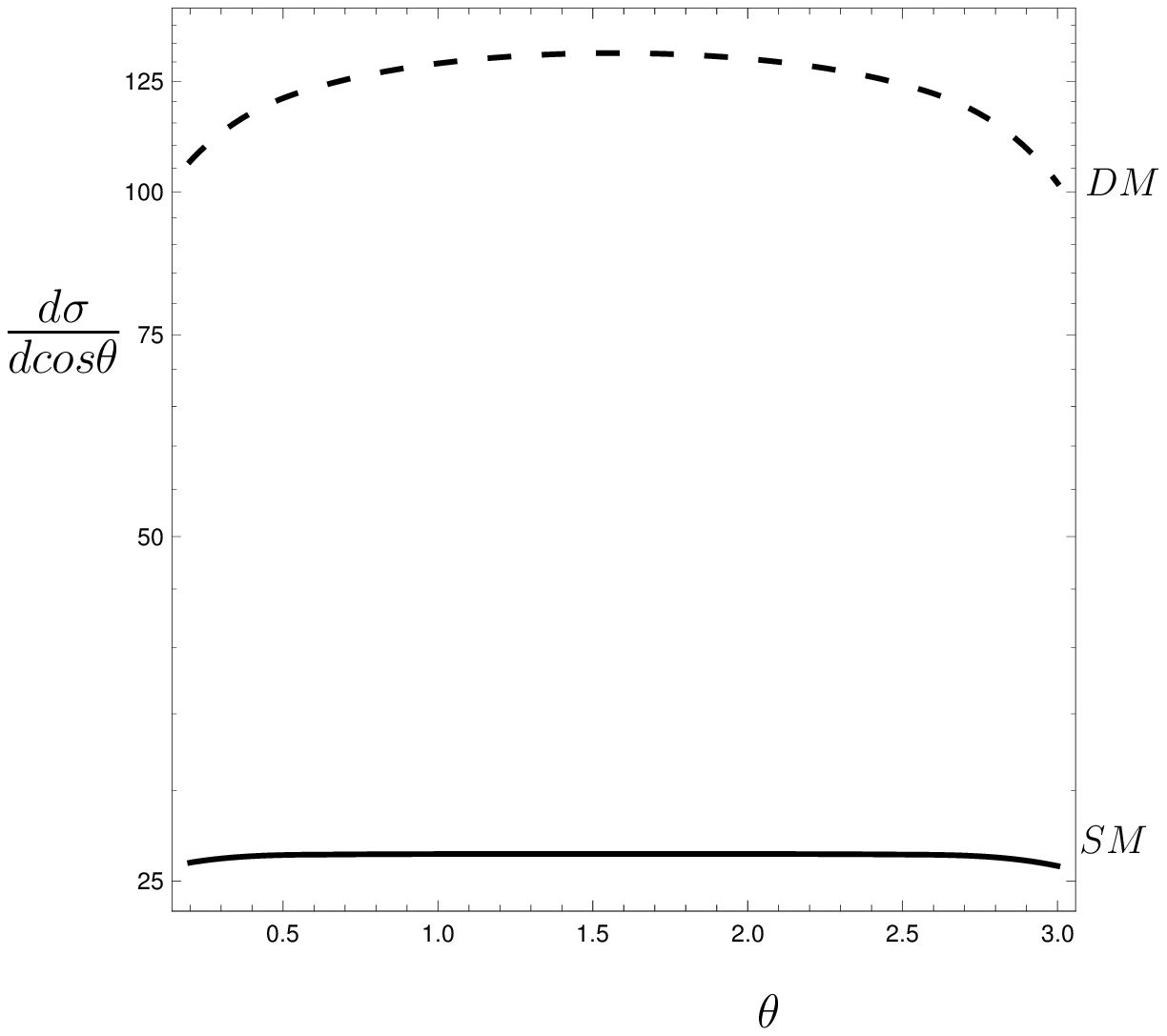 , height=20.cm}
\]\\
\vspace{-10cm}
\caption[1] {DM effect in $ZZ \to ZZ$; energy dependence in upper level
and angular dependence in lower level.}
\end{figure}

\clearpage

\begin{figure}[p]
\[
\epsfig{file=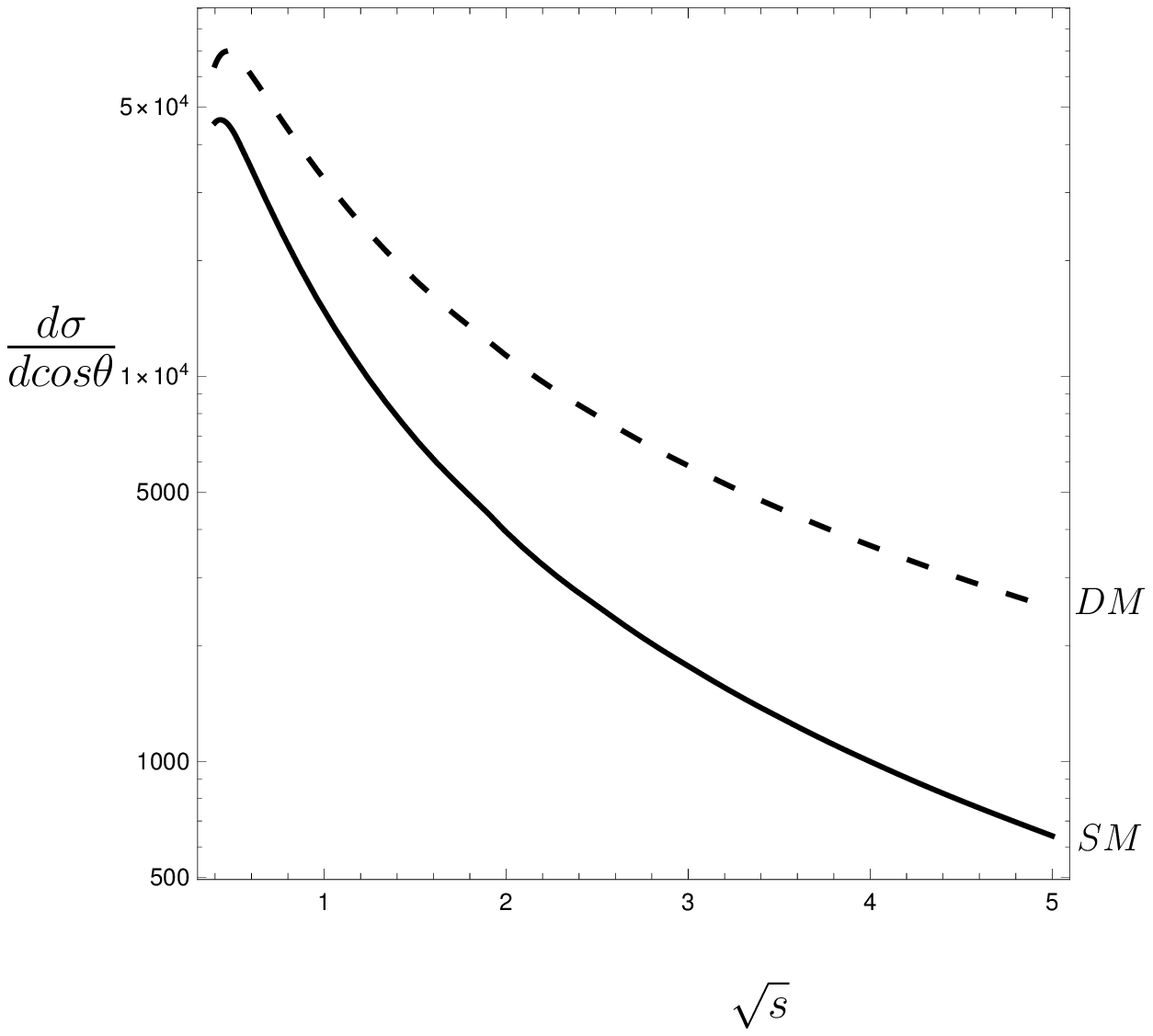 , height=20.cm}
\]\\
\vspace{-12cm}
\[
\epsfig{file=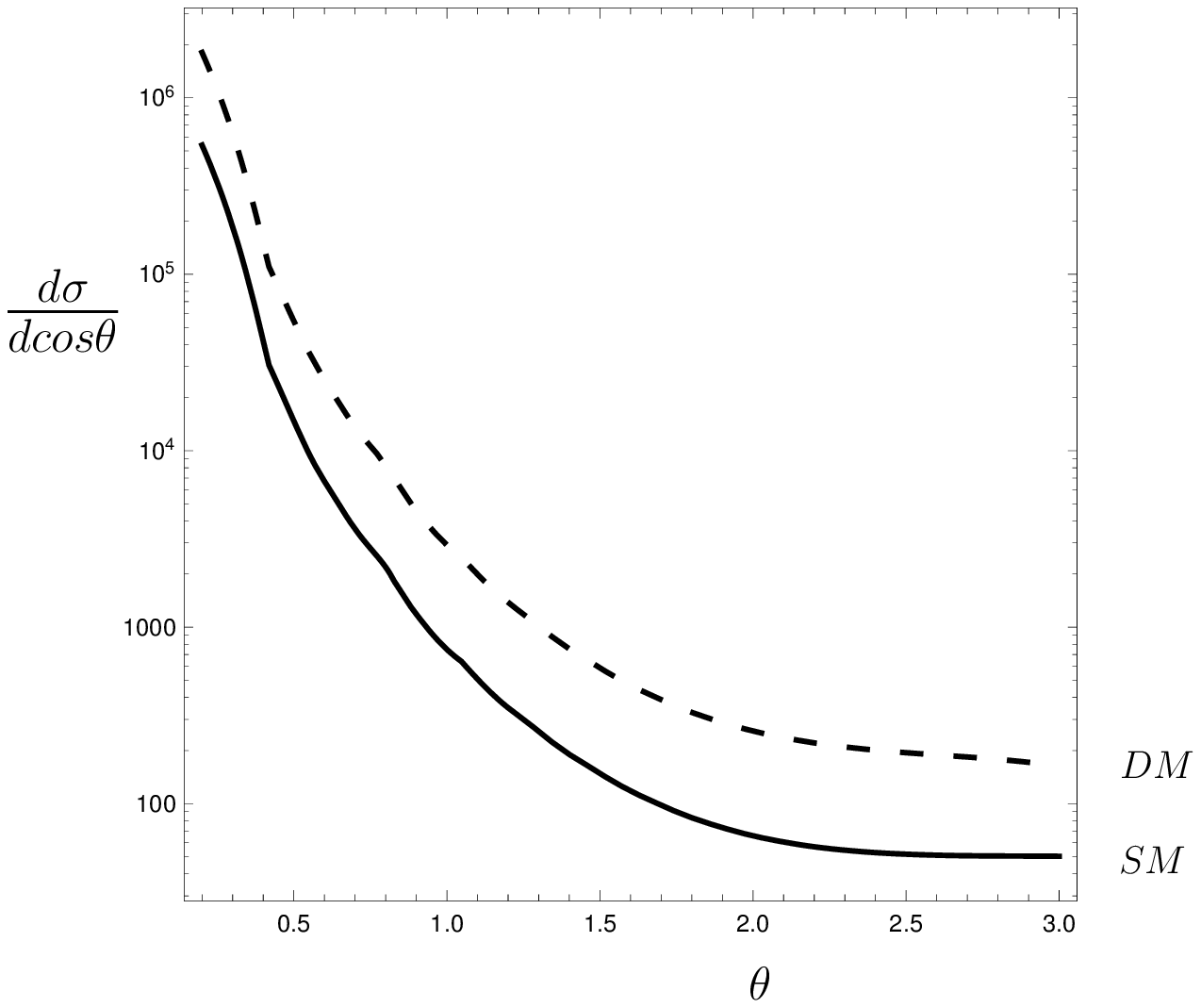 , height=20.cm}
\]\\
\vspace{-10cm}
\caption[1] {DM effect in $W^+W^- \to W^+W^-$; energy dependence in upper level
and angular dependence in lower level.}
\end{figure}

\clearpage

\begin{figure}[p]
\[
\epsfig{file=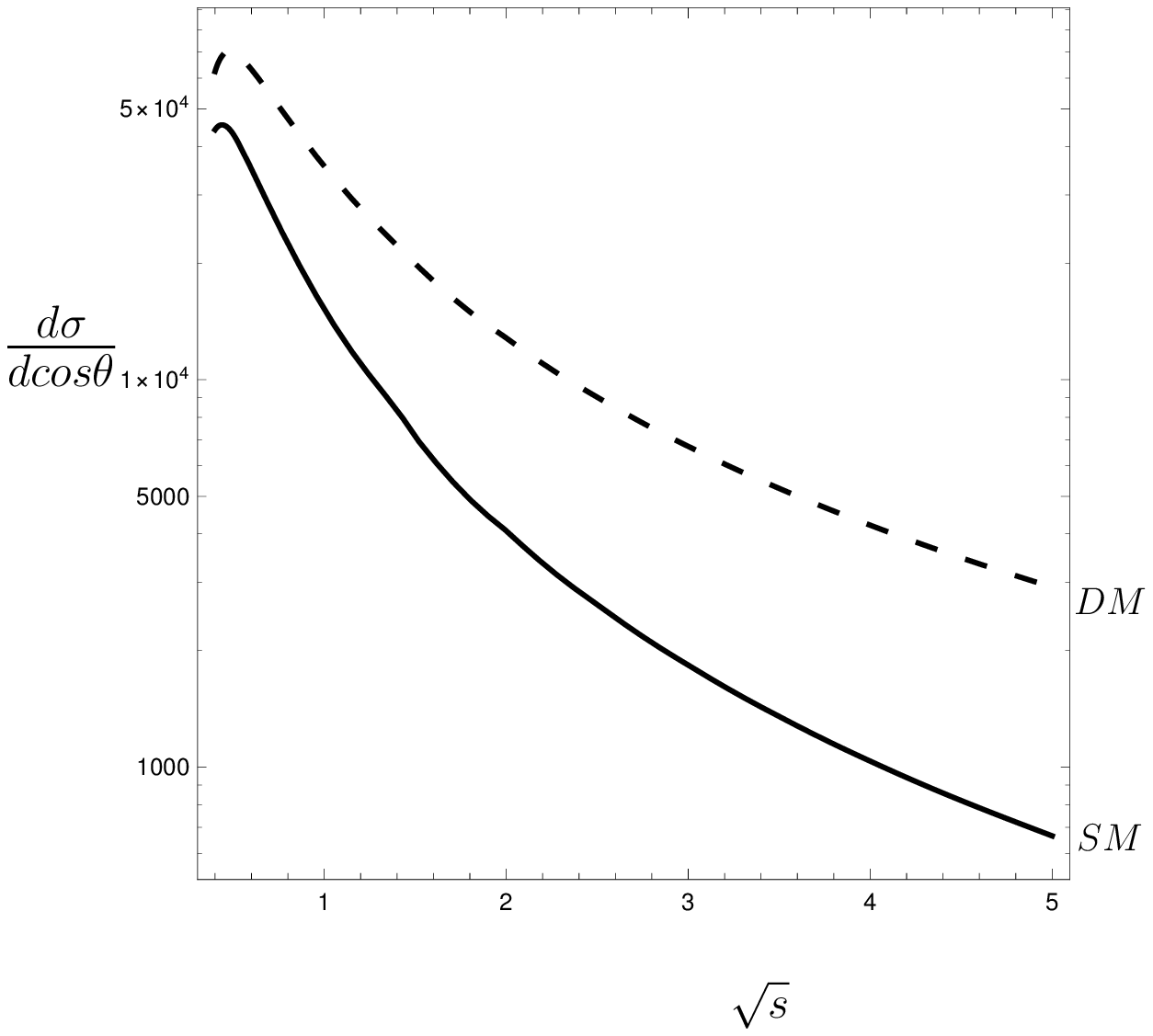 , height=20.cm}
\]\\
\vspace{-12cm}
\[
\epsfig{file=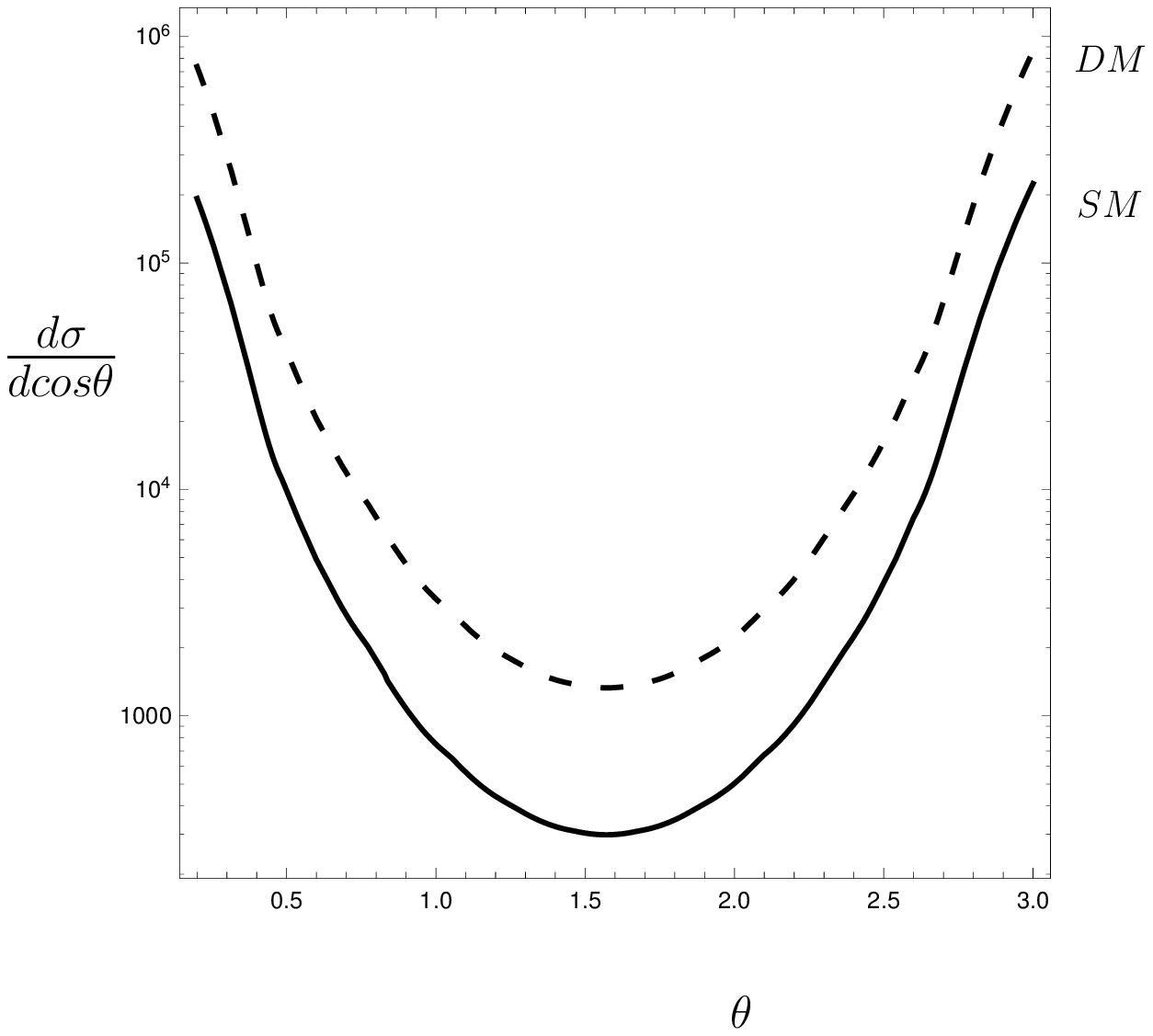 , height=20.cm}
\]\\
\vspace{-10cm}
\caption[1] {DM effect in $ZZ \to W^+W^-$; energy dependence in upper level
and angular dependence in lower level.}
\end{figure}

\clearpage

\begin{figure}[p]
\[
\epsfig{file=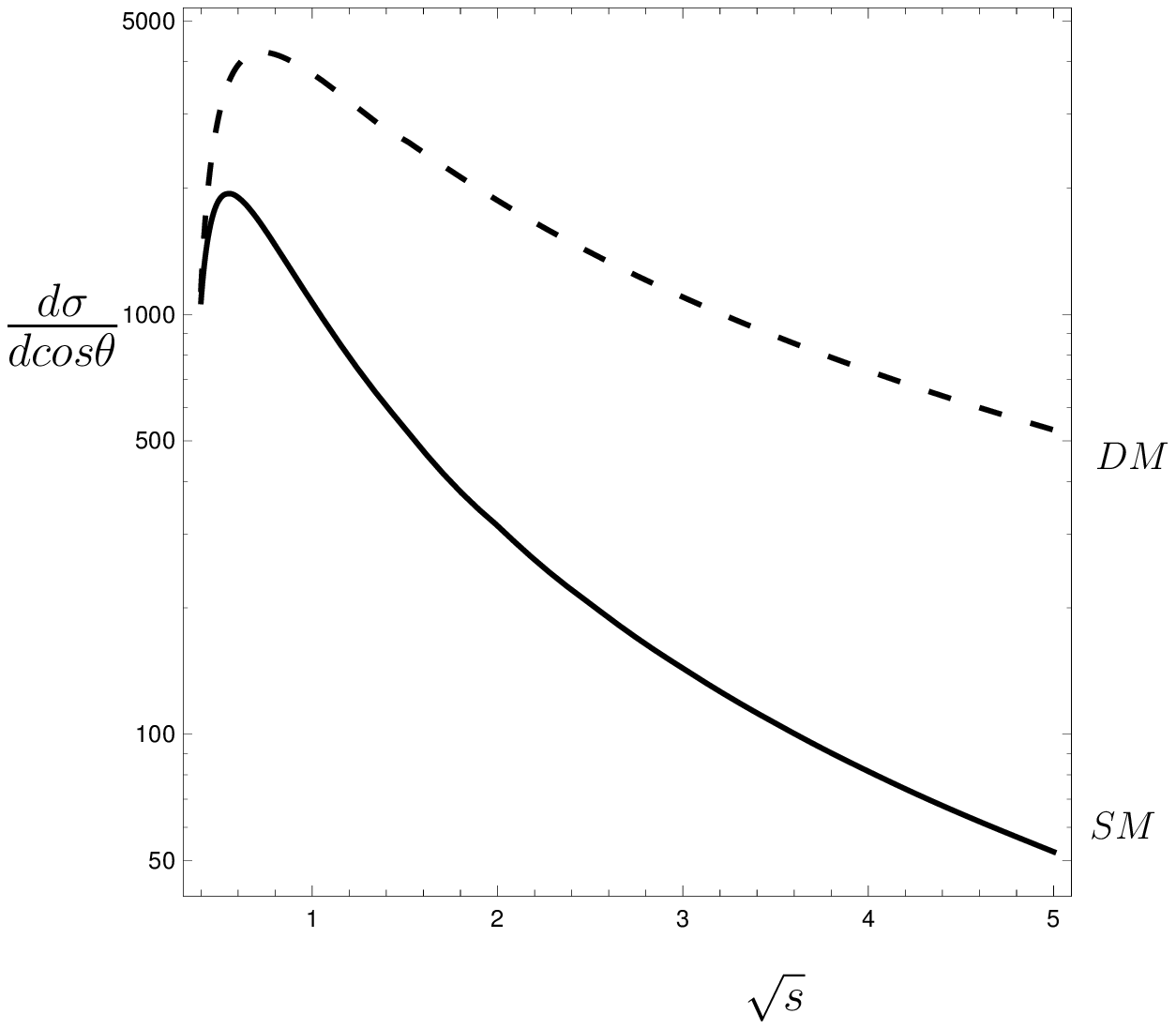 , height=20.cm}
\]\\
\vspace{-12cm}
\[
\epsfig{file=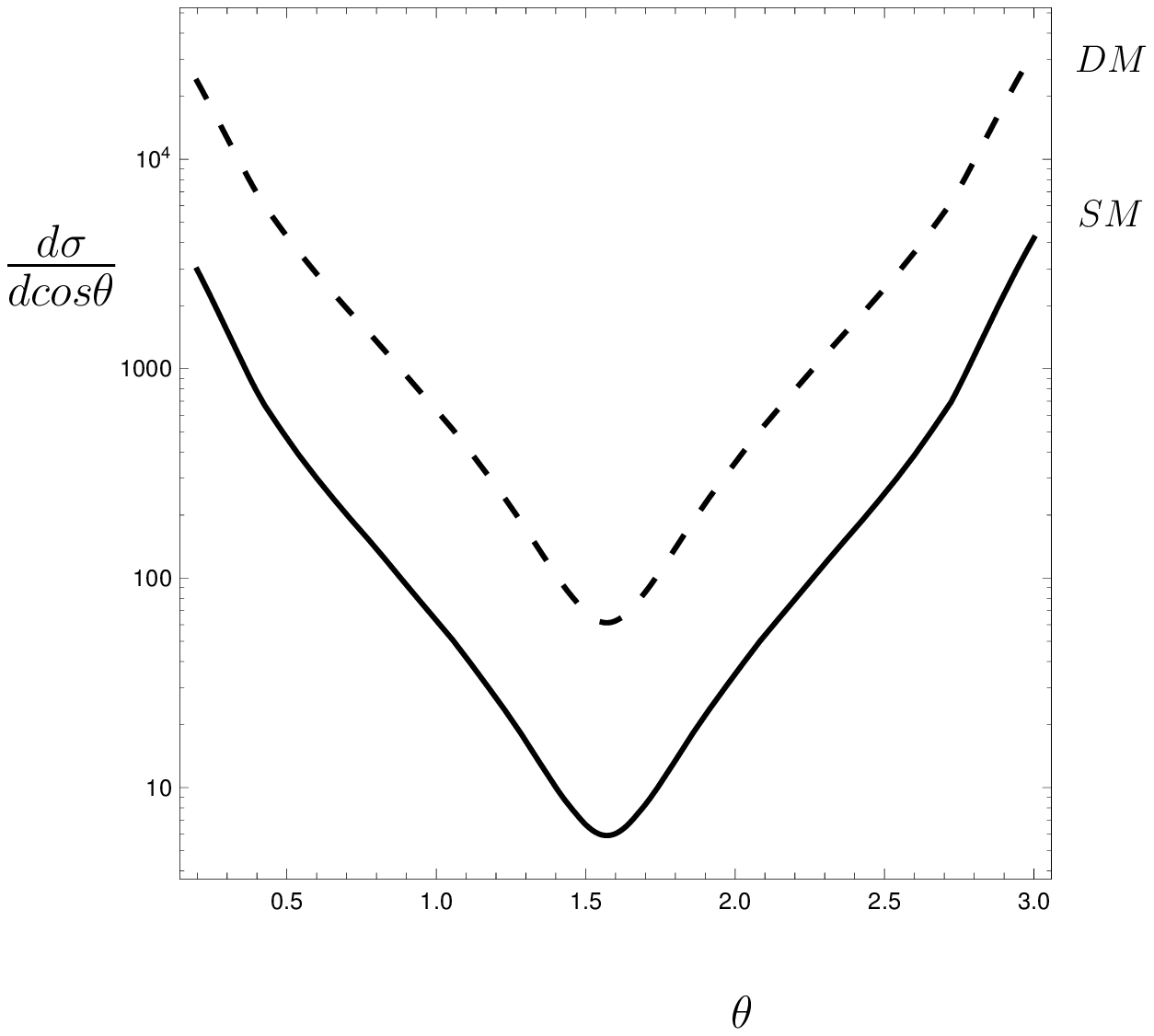 , height=20.cm}
\]\\
\vspace{-10cm}
\caption[1] {DM effect in $ZZ \to t\bar t$; energy dependence in upper level
and angular dependence in lower level.}
\end{figure}

\clearpage

\begin{figure}[p]
\[
\epsfig{file=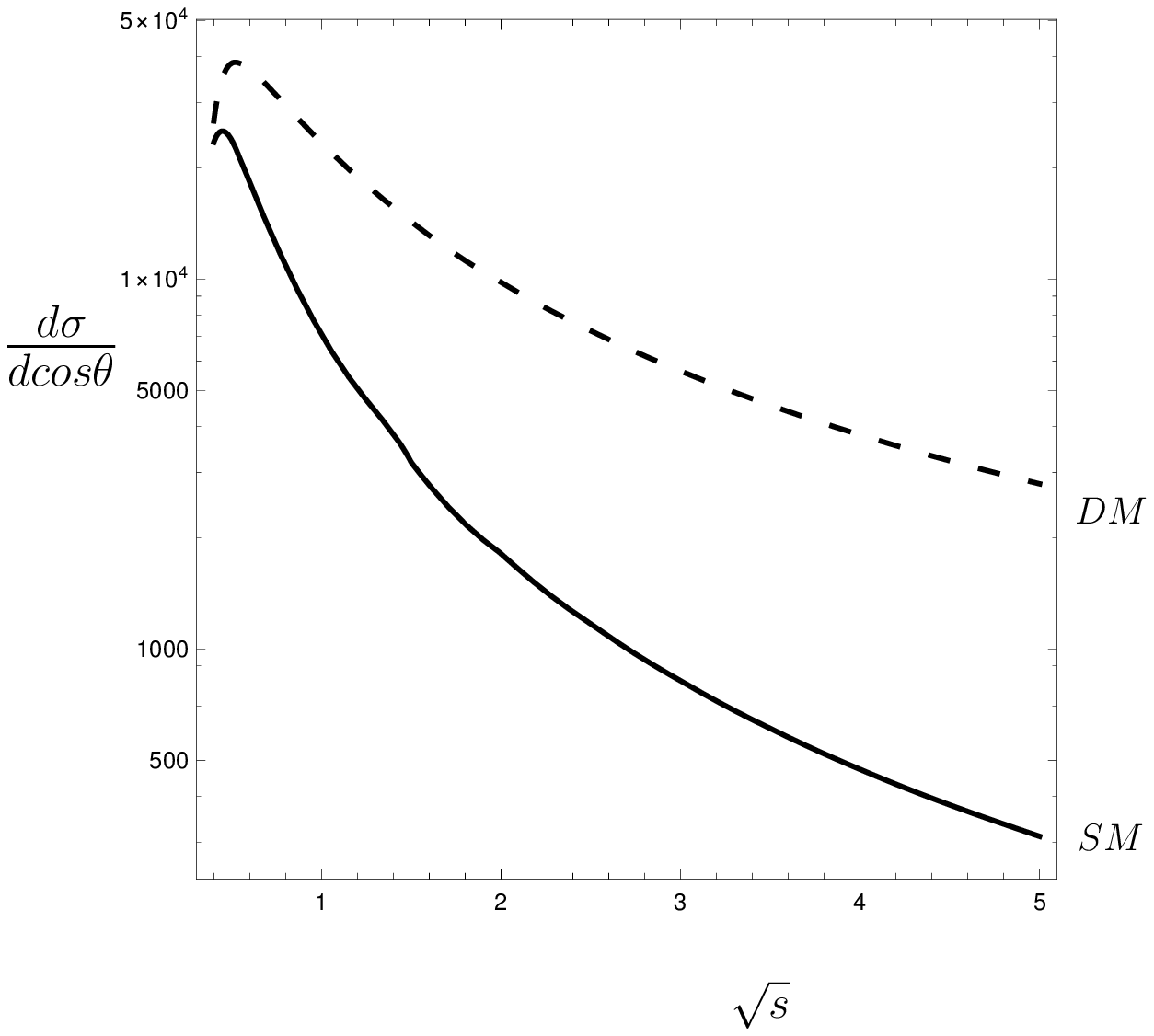 , height=20.cm}
\]\\
\vspace{-12cm}
\[
\epsfig{file=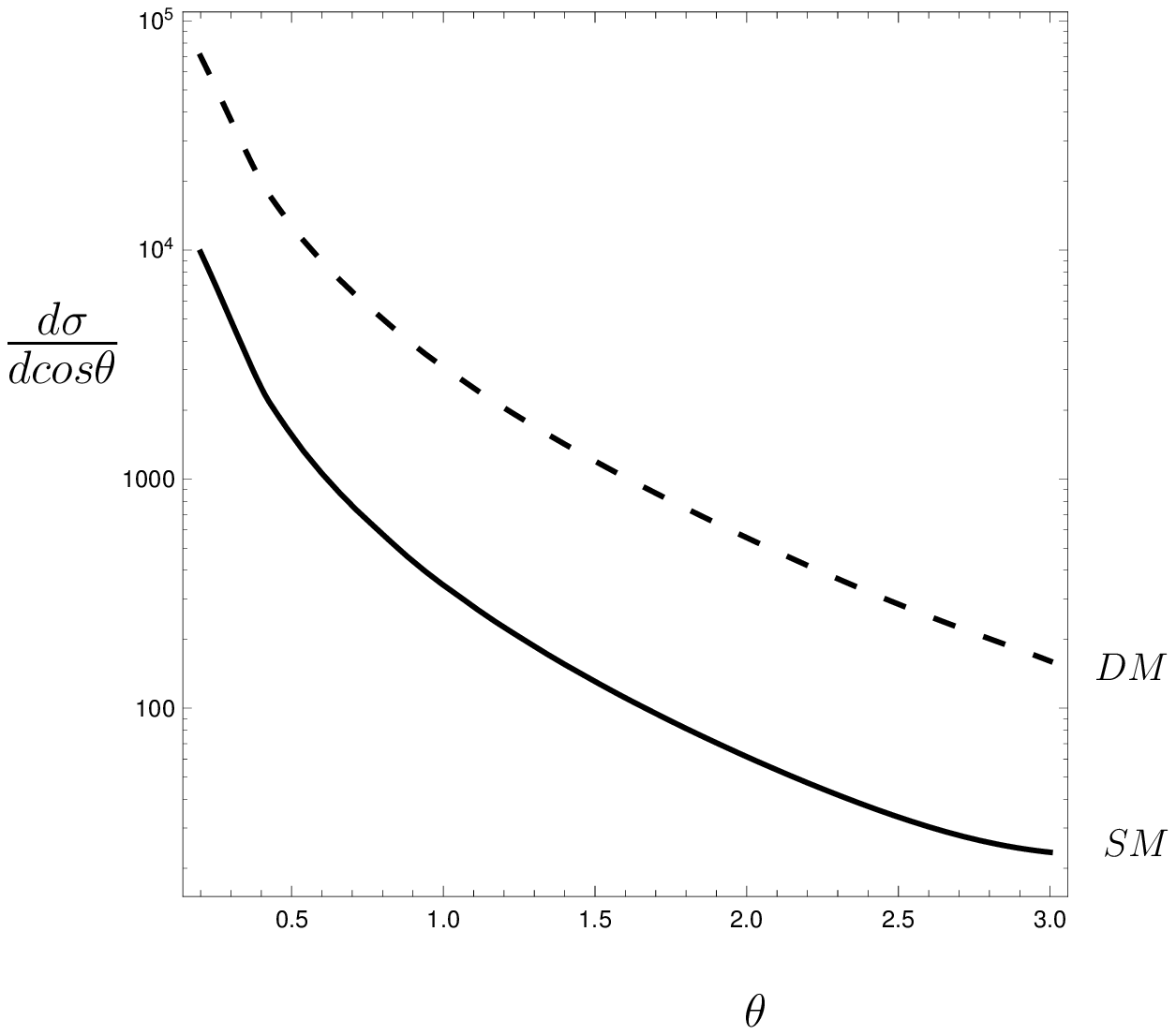 , height=20.cm}
\]\\
\vspace{-10cm}
\caption[1] {DM effect in $W^+W^- \to t\bar t$; energy dependence in upper level
and angular dependence in lower level.}
\end{figure}

\clearpage

\end{document}